\documentclass[a4paper]{article}

\usepackage{algorithm}
\usepackage{algorithmic}
\usepackage{INTERSPEECH2021}
\usepackage{amsfonts,subfigure,cite,epsfig,etoolbox,enumitem}
\usepackage{multirow}
\usepackage{amsmath}
\usepackage{amssymb}
\usepackage{ulem}
\usepackage{color}

\title{TeCANet: Temporal-Contextual Attention Network for Environment-Aware Speech Dereverberation}
\name{Helin Wang$^1$, Bo Wu$^2$, Lianwu Chen$^2$, Meng Yu$^3$, Jianwei Yu$^2$, Yong Xu$^3$, Shi-Xiong Zhang$^3$, Chao Weng$^2$, Dan Su$^2$, Dong Yu$^3$ \thanks{Part of this work performed when Helin Wang was interning at Tencent AI Lab.}}
\address{
  $^1$Peking University, Shenzhen, China\\
  $^2$Tencent AI Lab, Shenzhen, China\\
  $^3$Tencent AI Lab, Bellevue, WA, USA}
\email{wanghl15@pku.edu.cn, \{lambowu, lianwuchen, raymondmyu, tomasyu, lucayongxu, auszhang, cweng, dansu, dyu\}@tencent.com}

\begin{document}

\maketitle
\begin{abstract}
In this paper,
we exploit the effective way to leverage contextual information to improve the
speech dereverberation performance in real-world reverberant environments.
We propose a temporal-contextual attention approach on the deep neural network (DNN) for environment-aware speech dereverberation,
which can adaptively attend to the contextual information.
More specifically,
a FullBand based Temporal Attention approach (FTA) is proposed,
which models the correlations between the fullband information of the context frames.
In addition, considering the difference between the attenuation of high frequency bands and low frequency bands (high frequency bands attenuate faster than low frequency bands) in the room impulse response (RIR), 
we also propose a SubBand based Temporal Attention approach (STA).
In order to guide the network to be more aware of the reverberant environments,
we jointly optimize the dereverberation network and the reverberation time (RT60) estimator in a multi-task manner.
Our experimental results indicate that the proposed method outperforms our previously proposed reverberation-time-aware DNN and the learned attention weights are fully physical consistent. 
We also report a preliminary yet promising dereverberation 
and recognition experiment 
on real test data.

\end{abstract}

\section{Introduction}

When speech signals are obtained in an enclosed space by one or more microphones positioned at a distance from the talker, 
the observed signal consists of a superposition of many delayed and attenuated copies of the speech signal due to multiple reflections from the surrounding walls, ceilings, and floors \cite{naylor2010speech}.
As a result, intelligibility and quality of speech is degraded 
especially when the reverberation effects are severe \cite{neuman2010combined}.
The performances for automatic speech recognition (ASR) \cite{yoshioka2012making},
hearing aids and source localization are also severely affected.
Therefore, 
reducing the effect of reverberation is beneficial for the speech applications.

\begin{figure}[t]
  \centerline{\includegraphics[width=2.5in]{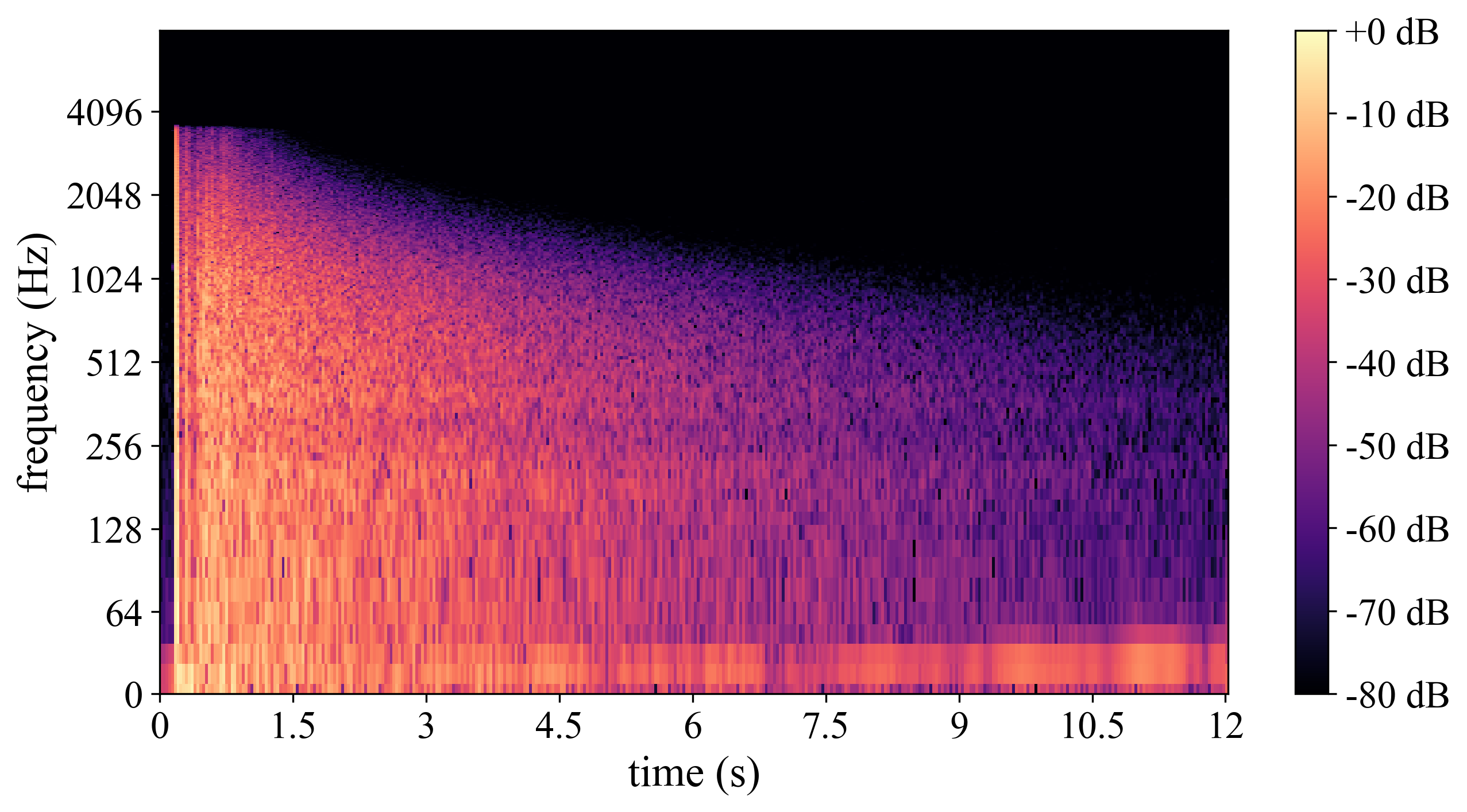}}
  \setlength{\abovecaptionskip}{0.cm}
  \caption{The spectrum of a real RIR.}  
  \label{fig1}
  \vspace{-0.5cm}
\end{figure}

To deal with the reverberation,
many dereverberation techniques have been developed in the past decades,
such as estimating an inverse filter of the room impulse response (RIR) \cite{wu2006two,neely1979invertibility},
and separating speech and reverberation via homomorphic transformation \cite{atal1974effectiveness,hermansky1994rasta}.
In recent years, 
deep neural networks (DNNs) \cite{hinton2006reducing} have been utilized in speech enhancement \cite{xu2014regression} and source separation \cite{huang2014deep},
and substantially outperformed conventional methods.
For speech dereverberation,
Han \textit{et al.} \cite{han2014learning}
firstly proposed to learn a spectral mapping from reverberant speech to anechoic speech with DNN.
Considering the importance of reverberation dependent parameters in supervised training,
Wu \textit{et al.} \cite{wu2016reverberation} developed a reverberation-time-aware DNN approach to suppress reverberation,
which investigated the effect of different contexts in a wide range of reverberation times (RT60s).
However,
the RT60-dependent context is manually chosen and a RT60 estimator is needed during the inference, which are quite unpractical in speech applications.
In order to capture long-term contextual information, 
Santos and Falk \cite{santos2018speech} proposed a context-aware recurrent neural network (RNN) method,
and Zhao \textit{et al.} \cite{zhao2020monaural} proposed a dereverberation algorithm using temporal convolutional networks.
Compared with speech denoising \cite{lu2013speech,tan2019real} which often benefits from incorporating long range contexts,
the correlations between the context frames are more crucial for speech dereverberation.
In a strong reverberant environment, the correlations between adjacent frames tend to be strong
and a long context is needed to capture adequate contextual information.
On the other hand, the long context may bring out redundant information
in a weak reverberant environment where the correlations between neighboring frames are weak.
In addition,
as shown in Figure~\ref{fig1},
different frequency bands attenuate diversely in a real RIR,
where the attenuation of high frequency bands is faster than that of low frequency bands \cite{naylor2010speech}.
Although several approaches had applied self attention networks \cite{zhao2020monaural,liu2020self,tan2020sagrnn} to explore the relevance among features at different time steps and frequency bands,
they ignored the physical relationships between the contextual information and reverberant environments and the different attenuation between the frequency bands,
which could be exploited to realize the better potential of neural networks.

In this paper,
we propose a novel attention-based approach for speech dereverberation,
which can adaptively attend to the contextual information by perceiving the reverberant environments.
More specifically,
a temporal attention module is first applied to the input features (\textit{i.e.} the log-power spectrum of reverberant speech)
to generate dynamic representations according to the correlations between frames,
and a dereveration network is then employed to 
learn the nonlinear mapping from the representations to the log-power spectrum of anechoic speech.
We use a small-footprint DNN model as the dereveration network 
due to its adaptation to practical applications.
Inspired by the distinct attenuation between the frequency bands in the RIR,
two types of approaches are proposed to obtain the attention weights,
including a FullBand based Temporal Attention approach (FTA) and a SubBand based Temporal Attention approach (STA).
The whole system is trained with the goal of both the dereverberation and the RT60 estimation,
so that it could be more aware of the reverberant conditions.
Our experiments are conducted on a public AISHELL-1 corpus \cite{bu2017aishell},
and the results show that our proposed method can significantly outperform our previous reverberation-time-aware methods \cite{wu2016reverberation}
and generalize well to the real RIRs.

\begin{figure}[t]
  \centerline{\includegraphics[width=3in]{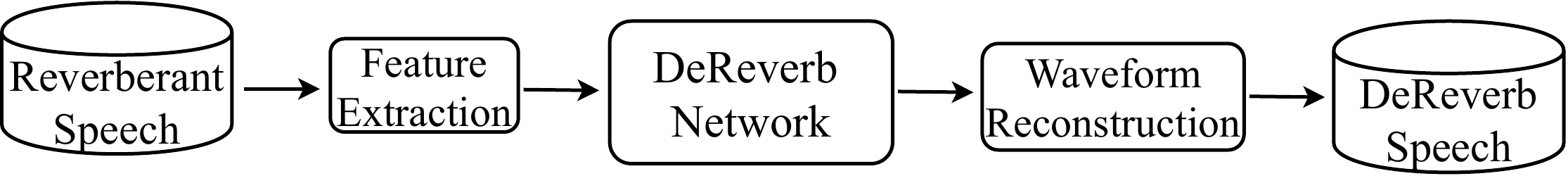}}
  \setlength{\abovecaptionskip}{0.2cm}
  \caption{Diagram of the DNN-based dereverberation system.}  
  \label{fig2}
\end{figure}

\begin{figure}[t] 
  \centering
  \subfigure[FTA]{\label{f31}
  \includegraphics[width=1.5in]{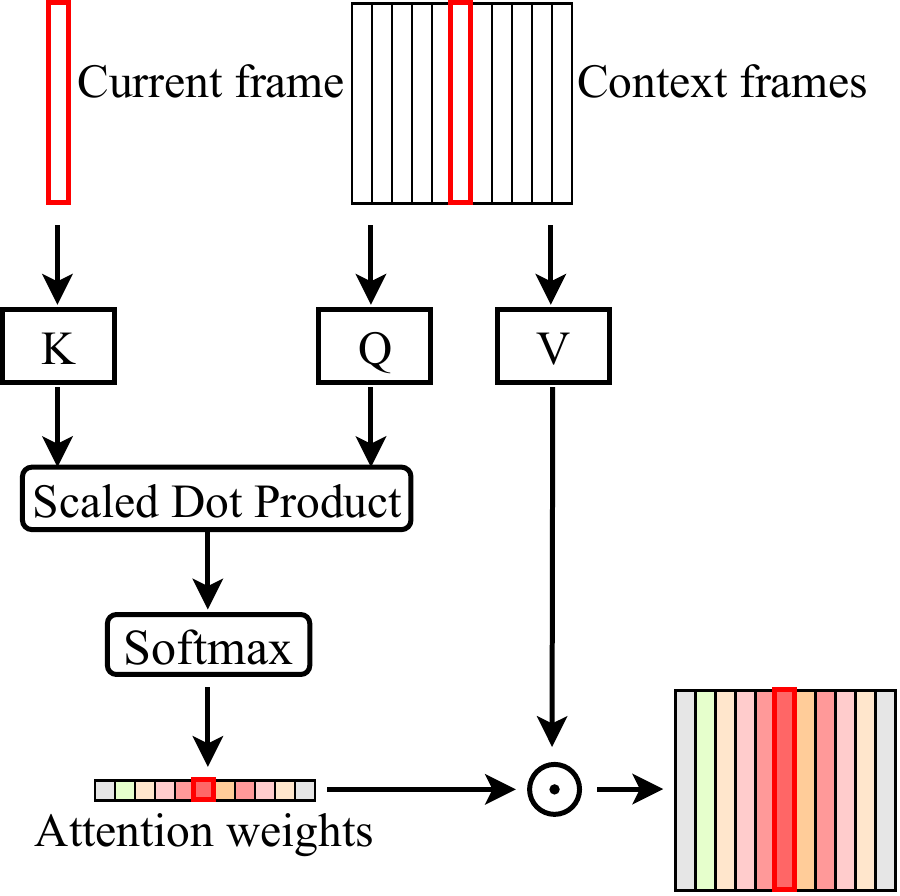}}
  \subfigure[STA]{\label{f32}
  \includegraphics[width=1.5in]{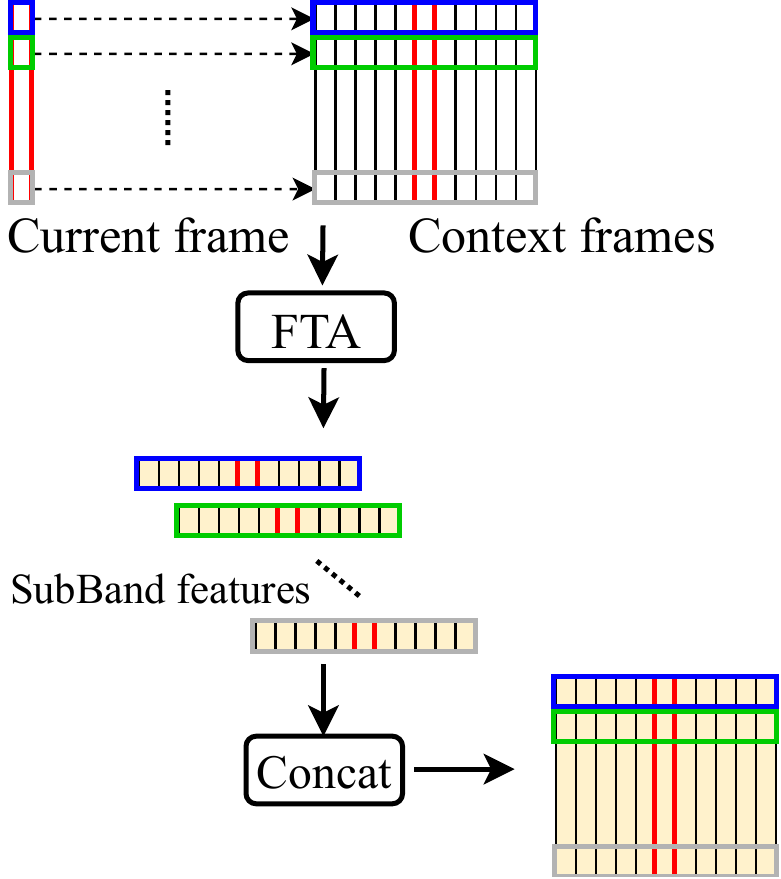}}
  \setlength{\abovecaptionskip}{0.cm}
  \caption{The illustration of FTA and STA.}
  \label{fig3}
  \vspace{-0.5cm}
\end{figure}

\section{Proposed method}
In this section,
we first describe the baseline DNN system. We refer the readers to \cite{wu2016reverberation} for details about the implementation of our previous work on reverberation-time-aware (RTA).
Then the proposed FullBand based Temporal Attention approach (FTA) and 
SubBand based Temporal Attention approach (STA) are detailed.
After that, we present our reverberation-environment-aware attention-based system, 
which implements both the dereverberation and the estimation of RT60s in the training stage.
 
\subsection{DNN-Based Speech Dereverberation System}
A block diagram of the DNN-based speech dereverberation system is illustrated in Figure~\ref{fig2},
which consists of a feature extraction module, a dereverberation network and a waveform reconstruction module.
Given the clean speech $s\left(t\right)$ and room impulse response function $h\left(t\right)$,
the reverberant speech $y\left(t\right)$ can be written as
\begin{equation}
  y\left(t\right)=s\left(t\right)*h\left(t\right)=x\left(t\right)+r\left(t\right)
  \label{eq1}
\end{equation}
where $*$ stands for the convolution operator,
$x\left(t\right)$ and $r\left(t\right)$ denote the anechoic speech (direct sound) and its reverberation, respectively.
Here, $x\left(t\right)$ is slightly different from $s\left(t\right)$ by a time shift and an energy decay caused by sound propagation through the direct path.
Our objective is to recover $x\left(t\right)$ from the corresponding reverberant observation $y\left(t\right)$.

Following \cite{wu2016reverberation},
the dereverberation task is to map the normalized log-power spectrum (LPS) of the reverberant speech to the normalized LPS anechoic speech.
Supposing the number of frames in the utterance is $T$ and the number of frequency channels is $F$, 
we use $\boldsymbol{Y}\in \mathbb{R}^{T\times F}$ to denote the extracted LPS features of the reverberant speech.
Similarly, the LPS of the anechoic speech can be denoted as $\boldsymbol{X}$, where $\boldsymbol{X}\in \mathbb{R}^{T\times F}$.
The dereverberation task is now formulated to be a sequence-to-sequence mapping problem,
\textit{i.e.},
$\{\boldsymbol{Y}(i)\} \rightarrow\{\boldsymbol{X}(i)\}, i=1,2,\ldots,T$.
Finally, the dereverberated waveform is reconstructed from the estimated LPS and the reverberant speech phase with an overlap-add method \cite{du2008speech}.

\begin{figure}[t]
  \centerline{\includegraphics[width=3in]{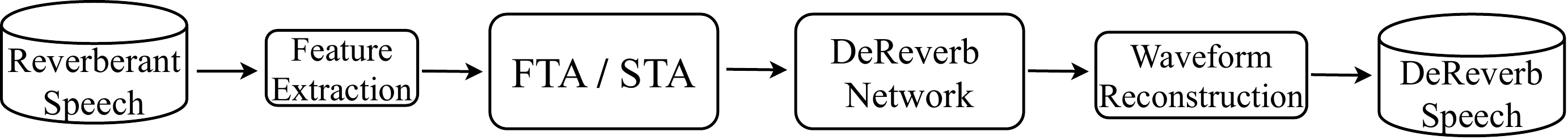}}
  \setlength{\abovecaptionskip}{0.2cm}
  \caption{Diagram of the attention-based system.}
  \label{fig4}
\end{figure}

\begin{figure}[t]
  \centerline{\includegraphics[width=3in]{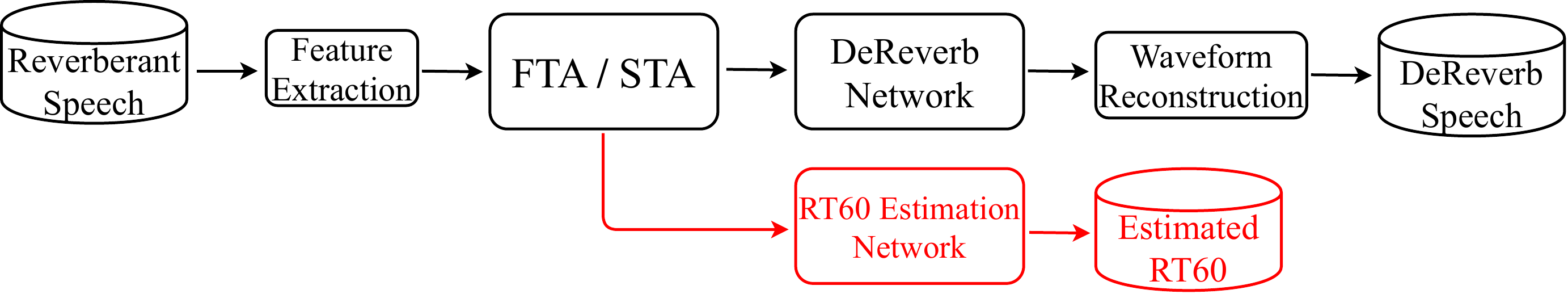}}
  \setlength{\abovecaptionskip}{0.2cm}
  \caption{Diagram of the reverberation-environment-aware attention-based system.}
  \label{fig5}
  \vspace{-0.5cm}
\end{figure}

\subsection{FullBand based Temporal Attention Approach}
In an enclosed space,
the received signal will be a collection of many delayed and attenuated copies of the original speech signal,
resulting in strong feature correlations at different time steps of the input sequence.
Reverberation information like reverberation time is embedded in these correlations,
and the correlations tend to be strong in a severe reverberant environment while weak in a weak reverberant situation.
In order to explore the input sequence to adapt to various reverberant environments,
we introduce the attention mechanism, which can dynamically attend to the relevant features.
Different from the self attention \cite{vaswani2017attention} that inputs the whole sequence,
our method utilizes the frames within a context window for each frame to calculate the dynamic features,
which greatly reduce the computational complexity.

\begin{table*}[t]
  \caption{PESQ, fwSegSNR, STOI and WER of different models. 
  Note that DNN-based models use \{1,3,5,7,9,11,13\}-frame expansion.}
  \label{tab1}
  \tiny
  \centering
  \begin{tabular}{|c|p{0.45cm}p{0.45cm}p{0.45cm}p{0.45cm}p{0.45cm}p{0.45cm}p{0.45cm}p{0.45cm}p{0.45cm}p{0.45cm}p{0.45cm}|c|c|c|c|c|}
    \hline
    \multirow{2}{*}{\textbf{Model}} & \multicolumn{12}{|c|}{\textbf{RT60 of simulated RIRs (s)}} & \multicolumn{4}{|c|}{\textbf{Real RIRs}} \\
    \cline{2-17}
     & 0$\sim$0.1 & 0.1$\sim$0.2& 0.2$\sim$0.3 & 0.3$\sim$0.4 & 0.4$\sim$0.5 & 0.5$\sim$0.6 & 0.6$\sim$0.7 & 0.7$\sim$0.8 & 0.8$\sim$0.9 & 0.9$\sim$1.0 & Avg.& WER & PESQ & fwSegSNR & STOI & WER \\
    \hline
    Rev & 3.56 & 3.45 & 3.10 & 2.83 & 2.65 & 2.52 & 2.40 & 2.28 & 2.20 & 2.10 & 2.67 & 33.8 & 3.05 &8.92&0.82&33.2\\
    \hline
    DNN-base-01 & \bf{3.44} & \bf{3.39}&3.18&2.97&2.82&2.70&2.60&2.48&2.40&2.29&2.80&-&-&-&-&-\\
    DNN-base-03 &3.39&3.35&\bf{3.25}&3.09&2.97&2.88&2.80&2.70&2.63&2.54&2.94&-&-&-&-&-\\
    DNN-base-05&3.38&3.35&3.25&\bf{3.13}&3.02&2.94&2.87&2.77&2.73&2.64&2.99&-&-&-&-&-\\
    DNN-base-07&3.37&3.34&3.24&3.11&\bf{3.04}&2.96&2.89&2.81&2.76&2.68&3.00&-&-&-&-&-\\
    DNN-base-09&3.37&3.34&3.24&3.13&3.03&\bf{2.97}&\bf{2.91}&\bf{2.83}&\bf{2.79}&2.70&\bf{3.02}&30.1&3.10&9.40&0.82&30.1\\
    DNN-base-11&3.35&3.32&3.22&3.11&3.02&2.95&2.89&2.82&2.78&\bf{2.70}&3.00&-&-&-&-&-\\
    DNN-base-13&3.34&3.31&3.21&3.10&3.01&2.94&2.88&2.80&2.76&2.69&2.99&-&-&-&-&-\\
    \hline
    DNN-RTA-EstRT60 \cite{wu2016reverberation}&3.44&3.39&3.26&3.13&3.04&2.97&2.92&2.83&2.80&2.71&3.04&28.6&3.08&9.39&0.82&32.1\\
    DNN-RTA-KnownRT60 \cite{wu2016reverberation}&3.45&3.39&3.29&3.14&3.04&2.99&2.95&2.87&2.82&2.78&3.06&-&-&-&-&-\\
    \hline
    \textbf{TeCANet-FTA (ours)}&3.40&3.37&3.27&3.15&3.06&2.99&2.93&2.86&2.82&2.74&3.04&27.7&3.14&9.56&0.83&28.1\\
    \textbf{TeCANet-STA (ours)}&3.47&3.44&3.32&3.19&3.09&3.02&2.97&2.89&2.85&2.78&3.09&25.7&3.18&9.88&0.84&28.1\\
    \textbf{TeCANet-FTA-EstRT60 (ours)}&3.46	&3.43	&3.30	&3.17	&3.07	&3.00	&2.94	&2.86	&2.83	&2.75	&3.06&25.3 &3.17 &9.73 &0.83&\bf{27.4}\\
    \textbf{TeCANet-STA-EstRT60 (ours)}&\bf{3.50}	&\bf{3.47}	&\bf{3.35}	&\bf{3.21}	&\bf{3.12}	&\bf{3.05}	&\bf{2.99}	&\bf{2.92}	&\bf{2.88}	&\bf{2.81}	&\bf{3.11}&\bf{23.7}&\bf{3.22}& \bf{9.98} & \bf{0.84}&28.4\\
    \hline
  \end{tabular}
\end{table*}

Figure~\ref{f31} shows the illustration of the proposed FullBand based Temporal Attention approach (FTA).
For the input feature of each time step $\boldsymbol{Y}\left(i\right)\in\mathbb{R}^{F}$,
a $c$-frame expansion is firstly applied \cite{wu2016reverberation} to obtain the context feature $\boldsymbol{C}\left(i\right)\in\mathbb{R}^{F\times c}$, which is then passed to three parallel layers: query, key, and value.
These layers map the input feature to a query $\boldsymbol{Q}\left(i\right)\in\mathbb{R}^{d_q\times c}$, 
a key $\boldsymbol{K}\left(i\right)\in\mathbb{R}^{d_q}$, 
and a value $\boldsymbol{V}\left(i\right)\in\mathbb{R}^{d_v\times c}$,
where $d_q$ and $d_v$ denote the dimension of the query and value, respectively.
\begin{gather}
  \label{eq2}
  \boldsymbol{Q}\left(i\right)=W_{Q}\boldsymbol{C}\left(i\right) \\
  \boldsymbol{K}\left(i\right)=W_{K}\boldsymbol{Y}\left(i\right) \\
  \boldsymbol{V}\left(i\right)=W_{V}\boldsymbol{C}\left(i\right)
\end{gather}
where $W_{Q}$, $W_{K}$ and $W_{V}$ denote the respective weight matrices.
The weight distribution on the context frames is computed based on the similarities between the query $\boldsymbol{Q}\left(i\right)$ and key $\boldsymbol{K}\left(i\right)$
by the scaled dot product.
\begin{equation}
  \boldsymbol{A}\left(i\right)=\operatorname{softmax}\left(\frac{\boldsymbol{Q}\left(i\right)^{T} \boldsymbol{K}\left(i\right)}{\sqrt{d_{\mathrm{q}}}}\right)
  \label{eq5}
\end{equation}
Multiplying the attention weights $\boldsymbol{A}\left(i\right) \in \mathbb{R}^{1\times c}$ by the value $\boldsymbol{V}\left(i\right)$, 
we get a weighted feature $\boldsymbol{Y^{\prime}}\left(i\right)\in \mathbb{R}^{d_v\times c}$.
\begin{equation}
  \boldsymbol{Y^{\prime}}\left(i\right)=\boldsymbol{V}\left(i\right)\odot \boldsymbol{A}\left(i\right)
  \label{eq6}
\end{equation}
where $\odot$ denotes the element-wise multiplication. 

\subsection{SubBand based Temporal Attention Approach}
FTA uses the fullband information of each frame to obtain the attention weights.
However, as shown in Figure~\ref{fig1},
the attenuation of a RIR is quite different between the frequency bands physically.
Thus, we should 
pay different attention to the contextual information in different frequency bands.
Figure~\ref{f32} shows our proposed SubBand based Temporal Attention approach (STA),
where we separate the fullband spectrum feature into a series of continuous subband spectrum features
and apply the FTA to each subband.
Let $\left\{\boldsymbol{Y}^1\left(i\right),\boldsymbol{Y}^2\left(i\right),\ldots,\boldsymbol{Y}^N\left(i\right) \right\}$
be the subband spectrum features of the $i$-th time step,
where $N$ denotes the number of subbands.
We can obtain the weighted features $\left\{\boldsymbol{Y}^{\prime^1}\left(i\right),\boldsymbol{Y}^{\prime^2}\left(i\right),\ldots,\boldsymbol{Y}^{\prime^N}\left(i\right) \right\}$
and the attention weights $\left\{\boldsymbol{A}^1\left(i\right),\boldsymbol{A}^2\left(i\right),\ldots,\boldsymbol{A}^N\left(i\right) \right\}$ with \eqref{eq2}-\eqref{eq6}.
Finally,
we merge the weighted features and the attention weights by concatenating them, respectively.

\subsection{The Reverberation-Environment-Aware System}
Figure~\ref{fig4} shows the attention-based system,
where FTA or STA is employed after the feature extraction module.
The weighted feature $\boldsymbol{Y^{\prime}}$ is concatenated and then fed to the dereverberation network.
In this case,
the attention weights are obtained by the similarity between the contexts.
In order to let the network be more aware of the reverberant environments,
we propose to jointly optimize the dereverberation and the estimation of RT60s, which is shown in Figure~\ref{fig5}.
More specifically,
the attention weights $\boldsymbol{A}$ are passed to another network to estimate the RT60 of each frame.
In the training stage,
the system implements both branches.
While in the inference stage,
only the dereverberation branch is used so that there are no extra parameters applied.
Let $\boldsymbol{\hat{X}}\in \mathbb{R}^{T\times F}$ and $\boldsymbol{\hat{Z}}\in \mathbb{R}^{T}$ be the output of the dereverberation network and the RT60 estimation network, respectively.
The mean squared error (MSE) is used as the loss function.
\begin{equation}
  \mathcal{L}(\boldsymbol{X}, \boldsymbol{\hat{X}}, \boldsymbol{Z}, \boldsymbol{\hat{Z}} ; \Theta)=\frac{1}{T\times F}\|\boldsymbol{X}-\boldsymbol{\hat{X}}\|_{2}^{2} + \frac{1}{T}\|\boldsymbol{Z}-\boldsymbol{\hat{Z}}\|_{2}^{2}
  \label{eq7}
\end{equation}
where $\|\bullet\|_{2}$ denotes the $l_2$ norm and $\Theta$ denotes the learnable parameters of the networks.
$\boldsymbol{X}\in \mathbb{R}^{T\times F}$ denotes the LPS of the anechoic speech
and $\boldsymbol{Z}\in \mathbb{R}^{T}$ denotes the ground truth of RT60 of the reverberant speech.

\section{Experiments}
\subsection{Datasets}
We simulated a reverberant version of public AISHELL-1 Mandarin corpus \cite{bu2017aishell}. 
There are $120,098$, $14,326$ and $7,176$ clean utterances to generate training data, validation data and test data, respectively. The classic image method \cite{lehmann2008prediction} is used to generate room impulse response (RIR) and reverberation time (RT60) ranges from $0.01$ to $1.0$ s. 
The room configuration (length-width-height) is randomly sampled from $3$-$3$-$2.5$ m to $10$-$8$-$6$ m. 
The microphone and speakers are at least $0.3$ m away from the wall. 
The distance between microphone array and speakers ranges from $0.5$ m to $10$ m.
All data is sampled at $16$ kHz.
In addition, we used real recored RIRs from the INTERSPEECH 2020 DNS Challenge\footnote{https://github.com/microsoft/DNS-Challenge} \cite{reddy2020interspeech},
and applied it to the test set.
Perceptual evaluation of speech quality (PESQ)\cite{recommendation2001perceptual},
frequency-weighted segmental signal-to-noise ratio (fwSegSNR) \cite{hu2007evaluation}
and short-time objective intelligibility (STOI) \cite{taal2011algorithm}
were used to evaluate the dereverberation results.
In addition, we
fed the dereverberated speech to an ASR system \cite{yu2018development}, and tested the word error rate (WER).

\begin{figure}[t]
  \centerline{\includegraphics[width=2.5in]{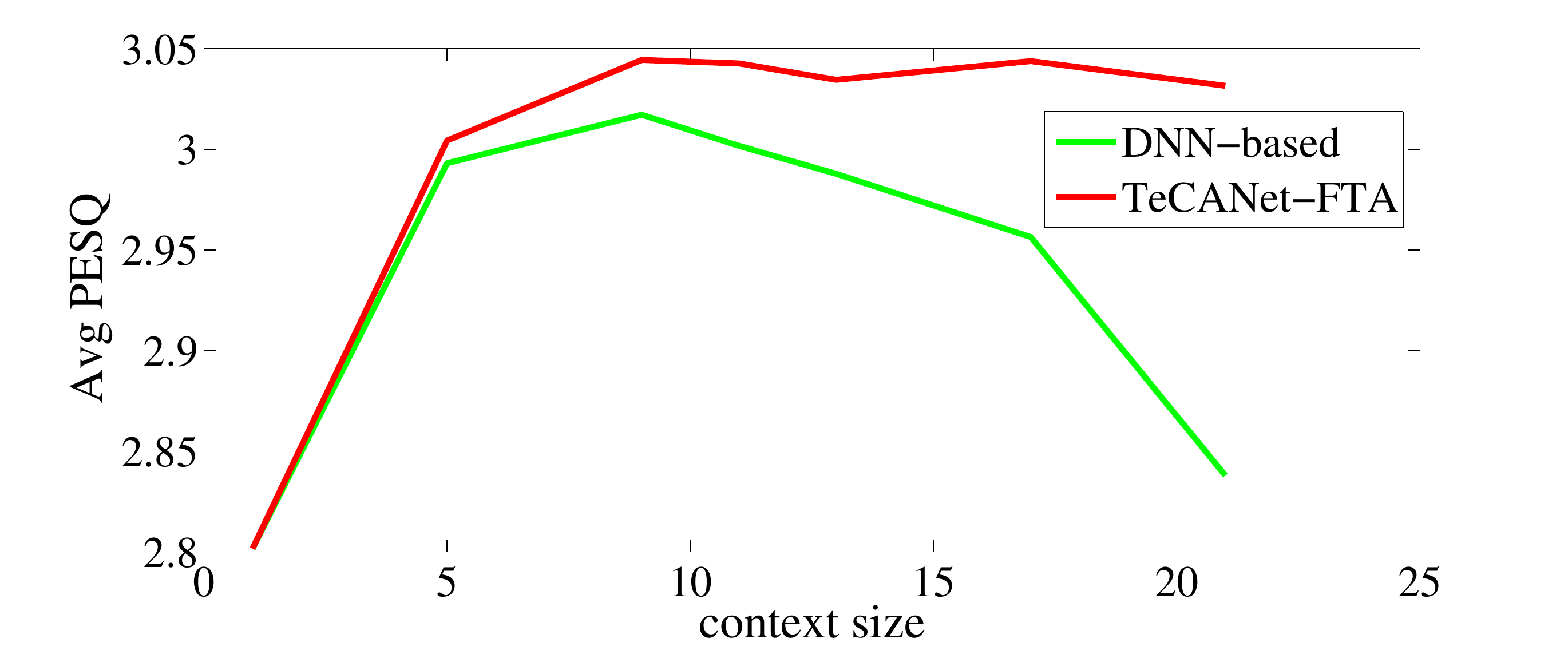}}
  \setlength{\abovecaptionskip}{0.1cm}
  \caption{Comparison of different context sizes.}  
  \label{fig6}
  \vspace{-0.3cm}
\end{figure}

\subsection{Setups}
For the input time-domain signal,
a $512$-point short-time Fourier transform (STFT) with a window size of $32$ ms and a window shift of $16$ ms is applied to each frame, 
followed by the logarithmic function,
which results in $257$ frequency bins.
Unless otherwise stated, the number of the expansion frames is $9$.
The dereverberation network is a feedforward neural network 
with $4$ hidden layers with $2048$, $2048$, $2048$ and $257$ hidden units.
The RT60 estimation network is a feedforward neural network with $3$ hidden layers with $64$, $64$ and $1$ hidden units,
followed by a sigmoid function.
For STA, the number of subbands is set to $8$. 
The Adam algorithm \cite{kingma2014adam} is employed as the optimizer.
The initial learning rate is $1 \times 10^{-4}$,
and decayed by $1 \times 10^{-5}$ if the loss does not reduce for continuous three epochs.
The mini-batch size is set to $16$, 
and training is terminated after $100$ epochs.

\begin{figure}[t]
  \centerline{\includegraphics[width=3.5in]{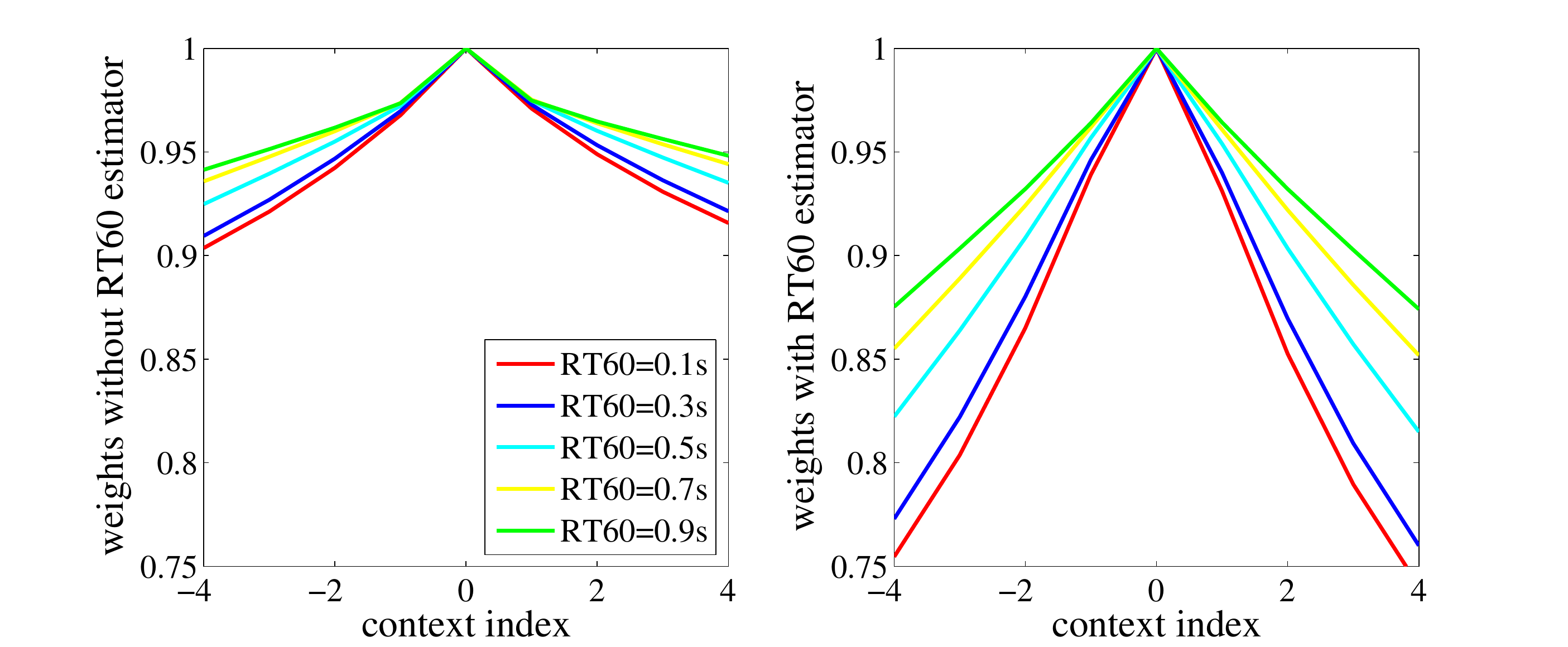}}
  \setlength{\abovecaptionskip}{0.1cm}
  \caption{Learned attention weights of FTA with/without RT60 Estimator.
  We calculated the mean of all the weights on the test set and normalized them by scaling the weight of current frame to $1$.
  Here, the context index $0$,$-n$ and $n$ stand for the current frame,
  the past $n$-th frame,
  and the future $n$-th frame, respectively.}   
  \label{fig7}
  \vspace{-0.5cm}
\end{figure}

\begin{figure}[t]
  \centerline{\includegraphics[width=3.5in]{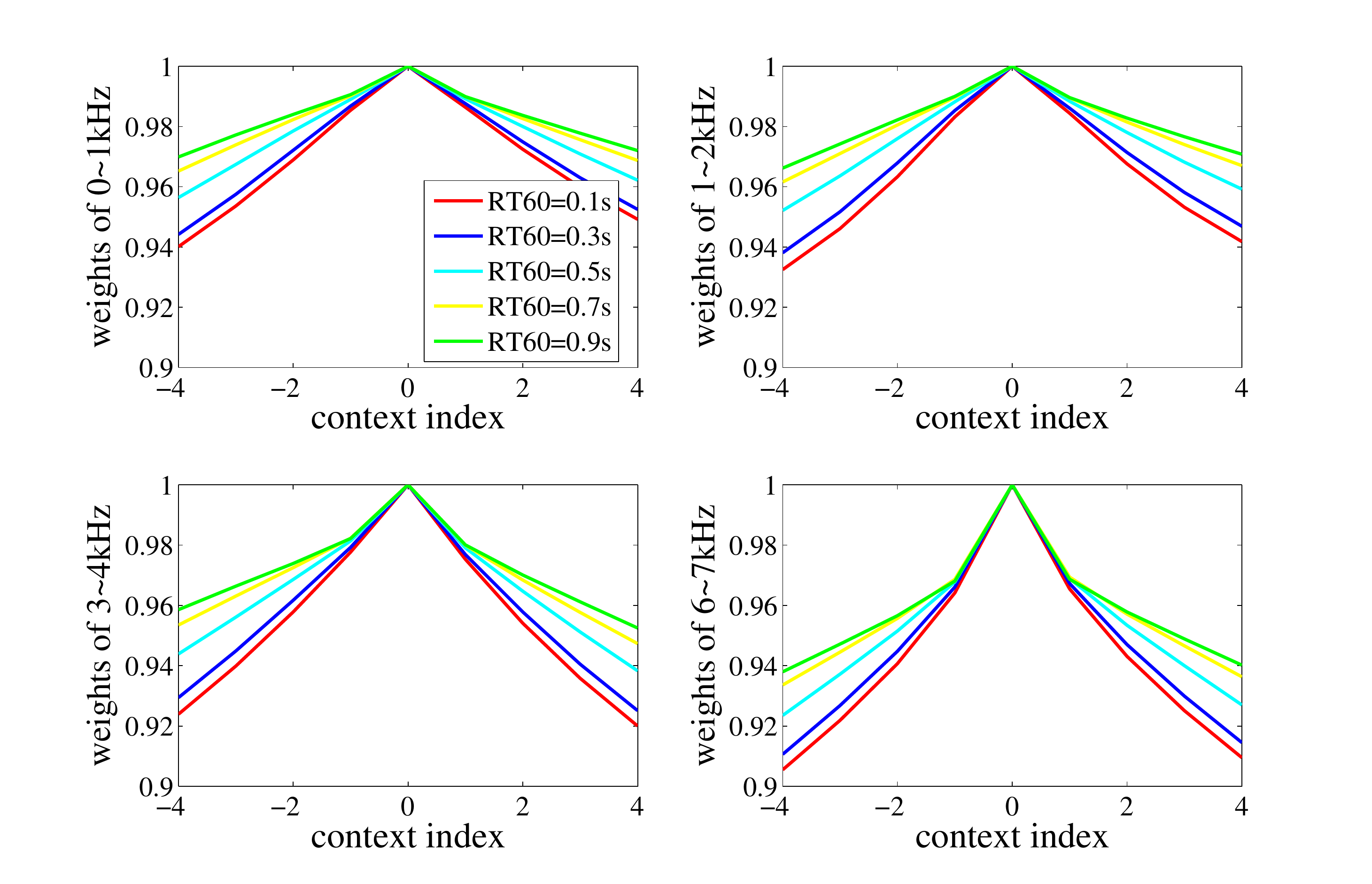}}
  \setlength{\abovecaptionskip}{0.cm}
  \caption{Learned attention weights of STA.
  We reported four of the frequency bands,
  including $0$-$1$ kHz, $1$-$2$ kHz, $3$-$4$ kHz and $6$-$7$ kHz.}  
  \label{fig8}
  \vspace{-0.5cm}
\end{figure}

\subsection{Results}
Table~\ref{tab1} shows the results of our proposed models and comparison models,
where we trained them on the training data with the simulated RIRs and tested them on the test data with the simulated RIRs and real RIRs.
DNN-based models with different frame expansions got different performances with the variation of the RT60s.
From DNN-base-01 to DNN-base-11,
we can see that the longer the RT60 was, 
the higher PESQ could be achieved with more context information.
But for a short RT60,
more context information could not improve performance,
and DNN-base-01 performed well,
especially when the RT60 is in the range of $0$$\sim$$0.2$ s.
DNN-base-13 could not beat others under all the RT60s
because too much redundant information was introduced.
Our previously proposed DNN-RTA-EstRT60 and DNN-RTA-KnownRT60 in \cite{wu2016reverberation} 
manually choose an optimal frame expansion according to the estimated and oracle RT60s during the inference stage, respectively,
which show better performance than the DNN-based models.
Thus, the main factor that influences the quality of enhanced speech
is not a long context but an appropriate context that is relevant to the reverberant environments. It should be noted that the RT60-dependent optimal context in DNN-RTA-EstRT60 is extracted from lots of DNN-based experiments and a high complexity RT60 estimator is needed in the inference stage.
Our proposed TeCANet-FTA obtained a comparable performance as DNN-RTA-KnownRT60,
owing to the FTA which models the correlations between the context frames.
Instead of manually choosing an optimal frame expansion, 
our method works by adaptively attending to the relevant contexts,
which adjusts to the real applications better.
As shown in Figure~\ref{fig6},
as the value of context size increases, 
the average PESQ on the test set of DNN-based model is firstly boosted, then achieves the peek when the context size is $9$, and finally dramatically drops because of lacking a mechanism that suppress redundant information.  
However, for TeCANet-FTA,
too much context information can hardly introduce the interference and
the performance tends to be stable when the context size is $9$ or more.

In addition, according to objective measures in both simulated and real data,
TeCANet-STA outperforms TeCANet-FTA,
which shows the effectiveness of paying different attention to different frequency bands.
In order to guide the networks to be more aware of the reverberant environments,
TeCANet-FTA-EstRT60 and TeCANet-STA-EstRT60 jointly optimize the dereverberation and the RT60 estimation in the training stage.
The results demonstrate that TeCANet-FTA-EstRT60 and TeCANet-STA-EstRT60 beat TeCANet-FTA and TeCANet-STA, respectively.
TeCANet-STA-EstRT60 achieves the highest PESQ ,STOI, and fwSegSNR among the models.
As for the ASR experiments,
the only difference is that TeCANet-FTA-EstRT60 obtains the lowest WER in real RIRs,
which may be caused by the gap between the dereverberation and recognition.
Although the subband information and the RT60 estimation seems to be less effective in this situation,
all our proposed attention-based models still outperform DNN-base-09, 
which shows good generalization to the real RIRs.

\subsection{Visualization Analysis}
In order to analyze the impacts of the attention module,
the learned attention weights are visualized.
Figure~\ref{fig7} shows the mean of attention weights of FTA on the test set.
It can be found out that the network learns to pay different attention to different context frames.
The weights of the frames near the current frame are larger than the weights of the frames far from the current frame,
and the current frame has the largest weight.
Meanwhile,
for different RT60s,
different attention weights are obtained for the same context index.
The longer the RT60 is, the stronger the correlations between the contexts are,
so that the attention weights of the frames far from the current frame are larger.
In addition, by jointly predicting RT60 value, which is also related to correlations between frames,
the network can be more aware of the reverberant environments.
We can see that the learned attention weights with RT60 estimator are more discriminative against the RT60s 
than the learned attention weights without RT60 estimator. 

Furthermore,
we visualized the learned attention weights of STA
which takes into account the difference between the frequency bands.
As shown in Figure~\ref{fig8}, 
the attention weights of the context frames for low frequency bands (\textit{e.g.} $0$-$1$ kHz) 
are quite larger than the ones for high frequency bands (\textit{e.g.} $6$-$7$ kHz),
which means that the network utilizes more contextual information for the low frequency bands.
This is exactly in accordance with the physical phenomena
that the attenuation of high frequency bands is quite faster than low frequency bands.

\section{Conclusions}
In this paper,
we have presented a reverberation-environment-aware attention-based approach for speech dereverberation,
which can adaptively attend to the contextual information.
We exploited the correlations between the context frames and the different attenuation between the frequency bands, and jointly optimized the RT60 estimation to let the network be more aware of the reverberant environments.
Experimental results and visualization analysis validated the advantages of our method.



\normalem
\bibliographystyle{IEEEtran}

\bibliography{mybib}

\begin{thebibliography}{10}
\providecommand{\url}[1]{#1}
\csname url@samestyle\endcsname
\providecommand{\newblock}{\relax}
\providecommand{\bibinfo}[2]{#2}
\providecommand{\BIBentrySTDinterwordspacing}{\spaceskip=0pt\relax}
\providecommand{\BIBentryALTinterwordstretchfactor}{4}
\providecommand{\BIBentryALTinterwordspacing}{\spaceskip=\fontdimen2\font plus
\BIBentryALTinterwordstretchfactor\fontdimen3\font minus
  \fontdimen4\font\relax}
\providecommand{\BIBforeignlanguage}[2]{{%
\expandafter\ifx\csname l@#1\endcsname\relax
\typeout{** WARNING: IEEEtran.bst: No hyphenation pattern has been}%
\typeout{** loaded for the language `#1'. Using the pattern for}%
\typeout{** the default language instead.}%
\else
\language=\csname l@#1\endcsname
\fi
#2}}
\providecommand{\BIBdecl}{\relax}
\BIBdecl

\bibitem{naylor2010speech}
P.~A. Naylor and N.~D. Gaubitch, \emph{Speech dereverberation}.\hskip 1em plus
  0.5em minus 0.4em\relax Springer Science \& Business Media, 2010.

\bibitem{neuman2010combined}
A.~C. Neuman, M.~Wroblewski, J.~Hajicek, and A.~Rubinstein, ``Combined effects
  of noise and reverberation on speech recognition performance of
  normal-hearing children and adults,'' \emph{Ear and Hearing}, vol.~31, no.~3,
  pp. 336--344, 2010.

\bibitem{yoshioka2012making}
T.~Yoshioka, A.~Sehr, M.~Delcroix, K.~Kinoshita, R.~Maas, T.~Nakatani, and
  W.~Kellermann, ``Making machines understand us in reverberant rooms:
  Robustness against reverberation for automatic speech recognition,''
  \emph{IEEE Signal Processing Magazine}, vol.~29, no.~6, pp. 114--126, 2012.

\bibitem{wu2006two}
M.~Wu and D.~Wang, ``A two-stage algorithm for one-microphone reverberant
  speech enhancement,'' \emph{IEEE Transactions on Audio, Speech, and Language
  Processing}, vol.~14, no.~3, pp. 774--784, 2006.

\bibitem{neely1979invertibility}
S.~T. Neely and J.~B. Allen, ``Invertibility of a room impulse response,''
  \emph{The Journal of the Acoustical Society of America}, vol.~66, no.~1, pp.
  165--169, 1979.

\bibitem{atal1974effectiveness}
B.~S. Atal, ``Effectiveness of linear prediction characteristics of the speech
  wave for automatic speaker identification and verification,'' \emph{the
  Journal of the Acoustical Society of America}, vol.~55, no.~6, pp.
  1304--1312, 1974.

\bibitem{hermansky1994rasta}
H.~Hermansky and N.~Morgan, ``Rasta processing of speech,'' \emph{IEEE
  Transactions on Speech and Audio Processing}, vol.~2, no.~4, pp. 578--589,
  1994.

\bibitem{hinton2006reducing}
G.~E. Hinton and R.~R. Salakhutdinov, ``Reducing the dimensionality of data
  with neural networks,'' \emph{Science}, vol. 313, no. 5786, pp. 504--507,
  2006.

\bibitem{xu2014regression}
Y.~Xu, J.~Du, L.-R. Dai, and C.-H. Lee, ``A regression approach to speech
  enhancement based on deep neural networks,'' \emph{IEEE/ACM Transactions on
  Audio, Speech, and Language Processing}, vol.~23, no.~1, pp. 7--19, 2014.

\bibitem{huang2014deep}
P.-S. Huang, M.~Kim, M.~Hasegawa-Johnson, and P.~Smaragdis, ``Deep learning for
  monaural speech separation,'' in \emph{IEEE International Conference on
  Acoustics, Speech and Signal Processing (ICASSP)}.\hskip 1em plus 0.5em minus
  0.4em\relax IEEE, 2014, pp. 1562--1566.

\bibitem{han2014learning}
K.~Han, Y.~Wang, and D.~Wang, ``Learning spectral mapping for speech
  dereverberation,'' in \emph{IEEE International Conference on Acoustics,
  Speech and Signal Processing (ICASSP)}.\hskip 1em plus 0.5em minus
  0.4em\relax IEEE, 2014, pp. 4628--4632.

\bibitem{wu2016reverberation}
B.~Wu, K.~Li, M.~Yang, and C.-H. Lee, ``A reverberation-time-aware approach to
  speech dereverberation based on deep neural networks,'' \emph{IEEE/ACM
  Transactions on Audio, Speech, and Language Processing}, vol.~25, no.~1, pp.
  102--111, 2016.

\bibitem{santos2018speech}
J.~F. Santos and T.~H. Falk, ``Speech dereverberation with context-aware
  recurrent neural networks,'' \emph{IEEE/ACM Transactions on Audio, Speech,
  and Language Processing}, vol.~26, no.~7, pp. 1236--1246, 2018.

\bibitem{zhao2020monaural}
Y.~Zhao, D.~Wang, B.~Xu, and T.~Zhang, ``Monaural speech dereverberation using
  temporal convolutional networks with self attention,'' \emph{IEEE/ACM
  Transactions on Audio, Speech, and Language Processing}, vol.~28, pp.
  1598--1607, 2020.

\bibitem{lu2013speech}
X.~Lu, Y.~Tsao, S.~Matsuda, and C.~Hori, ``Speech enhancement based on deep
  denoising autoencoder.'' in \emph{Proc. Interspeech}, vol. 2013, 2013, pp.
  436--440.

\bibitem{tan2019real}
K.~Tan, X.~Zhang, and D.~Wang, ``Real-time speech enhancement using an
  efficient convolutional recurrent network for dual-microphone mobile phones
  in close-talk scenarios,'' in \emph{IEEE International Conference on
  Acoustics, Speech and Signal Processing (ICASSP)}.\hskip 1em plus 0.5em minus
  0.4em\relax IEEE, 2019, pp. 5751--5755.

\bibitem{liu2020self}
C.~Liu and Y.~Sato, ``Self-attention for multi-channel speech separation in
  noisy and reverberant environments,'' in \emph{2020 Asia-Pacific Signal and
  Information Processing Association Annual Summit and Conference (APSIPA
  ASC)}.\hskip 1em plus 0.5em minus 0.4em\relax IEEE, 2020, pp. 794--799.

\bibitem{tan2020sagrnn}
K.~Tan, B.~Xu, A.~Kumar, E.~Nachmani, and Y.~Adi, ``{SAGRNN}: Self-attentive
  gated rnn for binaural speaker separation with interaural cue preservation,''
  \emph{IEEE Signal Processing Letters}, 2020.

\bibitem{bu2017aishell}
H.~Bu, J.~Du, X.~Na, B.~Wu, and H.~Zheng, ``Aishell-1: An open-source mandarin
  speech corpus and a speech recognition baseline,'' in \emph{20th Conference
  of the Oriental Chapter of the International Coordinating Committee on Speech
  Databases and Speech I/O Systems and Assessment (O-COCOSDA)}.\hskip 1em plus
  0.5em minus 0.4em\relax IEEE, 2017, pp. 1--5.

\bibitem{du2008speech}
J.~Du and Q.~Huo, ``A speech enhancement approach using piecewise linear
  approximation of an explicit model of environmental distortions,'' in
  \emph{Ninth Annual Conference of the International Speech Communication
  Association}, 2008.

\bibitem{vaswani2017attention}
A.~Vaswani, N.~Shazeer, N.~Parmar, J.~Uszkoreit, L.~Jones, A.~N. Gomez,
  {\L}.~Kaiser, and I.~Polosukhin, ``Attention is all you need,'' in
  \emph{Proceedings of the 31st International Conference on Neural Information
  Processing Systems}, 2017, pp. 6000--6010.

\bibitem{lehmann2008prediction}
E.~A. Lehmann and A.~M. Johansson, ``Prediction of energy decay in room impulse
  responses simulated with an image-source model,'' \emph{The Journal of the
  Acoustical Society of America}, vol. 124, no.~1, pp. 269--277, 2008.

\bibitem{reddy2020interspeech}
C.~K. Reddy, V.~Gopal, R.~Cutler, E.~Beyrami, R.~Cheng, H.~Dubey,
  S.~Matusevych, R.~Aichner, A.~Aazami, S.~Braun \emph{et~al.}, ``{The
  INTERSPEECH 2020 Deep Noise Suppression Challenge: Datasets, Subjective
  Testing Framework, and Challenge Results},'' \emph{Proc. Interspeech}, pp.
  2492--2496, 2020.

\bibitem{recommendation2001perceptual}
I.-T. Recommendation, ``Perceptual evaluation of speech quality ({PESQ}): An
  objective method for end-to-end speech quality assessment of narrow-band
  telephone networks and speech codecs,'' \emph{Rec. ITU-T P. 862}, 2001.

\bibitem{hu2007evaluation}
Y.~Hu and P.~C. Loizou, ``Evaluation of objective quality measures for speech
  enhancement,'' \emph{IEEE Transactions on audio, speech, and language
  processing}, vol.~16, no.~1, pp. 229--238, 2007.

\bibitem{taal2011algorithm}
C.~H. Taal, R.~C. Hendriks, R.~Heusdens, and J.~Jensen, ``An algorithm for
  intelligibility prediction of time--frequency weighted noisy speech,''
  \emph{IEEE Transactions on Audio, Speech, and Language Processing}, vol.~19,
  no.~7, pp. 2125--2136, 2011.

\bibitem{yu2018development}
J.~Yu, X.~Xie, S.~Liu, S.~Hu, M.~W. Lam, X.~Wu, K.~H. Wong, X.~Liu, and
  H.~Meng, ``{Development of the CUHK Dysarthric Speech Recognition System for
  the UA Speech Corpus},'' \emph{Proc. Interspeech}, pp. 2938--2942, 2018.

\bibitem{kingma2014adam}
D.~P. Kingma and J.~Ba, ``Adam: A method for stochastic optimization,'' in
  \emph{3rd International Conference on Learning Representations (ICLR)}, 2015.

\end{thebibliography}


\end{document}